\newcommand{\gradi}{\ifmmode^\circ\else$^\circ$\fi}

\documentstyle[epsfig]{elsart}
\begin{document}

\begin{frontmatter}

\title{The GRAAL high resolution BGO calorimeter and its 
energy calibration and monitoring system}

\author{F. Ghio, B. Girolami}
\address{Istituto Superiore di Sanit\`{a} and INFN Sezione Sanit\`{a},
Viale Regina Elena 299, I-00161, Roma, Italy}
\author{M. Capogni, L. Casano, L. Ciciani, A. D'Angelo, R. Di Salvo,}
\author{L. Hu, D. Moricciani, L. Nicoletti, G. Nobili, C. Schaerf} 
\address{INFN Sezione Roma2 and Universit\`{a} di Roma 'Tor Vergata',
Via della Ricerca Scientifica 1, I-00133 Roma, Italy}
\author{P. Levi Sandri}
\address{INFN Laboratori Nazionali di Frascati,
Via E. Fermi 40, PO Box 13, 
I-00044 Frascati, Italy}
\author{M. Castoldi, A. Zucchiatti}
\address{INFN Sezione di Genova and Universit\`{a} di Genova,
Via Dodecaneso 33, 
I-16146 Genova, Italy} 
\author{V. Bellini} 
\address{INFN Laboratori Nazionali del Sud and Universit\`{a} di Catania, 
Corso Italia 57,
I-95129 Catania, Italy} 

\begin{abstract}
We describe the electromagnetic calorimeter built for the GRAAL 
apparatus at the ESRF. Its monitoring system
is presented in detail. Results from tests and
the performance obtained during the first GRAAL
experiments are given. The energy calibration accuracy and stability
reached is a small fraction of the intrinsic detector
resolution.
\end{abstract}
\end{frontmatter}

\section{Introduction}

The GRAAL beam line facility \cite{report}
currently in operation at the ESRF in Grenoble,
is the first source of high intensity and
completely polarized $\gamma$ rays in the
energy range 0.4$\div$1.5 GeV.
This project has been realized, with the prevailing
support of the Istituto Nazionale di Fisica
Nucleare (INFN), to study polarization observables
in photoproduction reactions including strangeness.
\par The GRAAL apparatus (see Fig. 1),
consists of a high resolution and large solid angle
BGO electromagnetic calorimeter combined with
multiwire proportional chambers (MWPC) that
covers a solid angle range of almost 4$\pi$.
Particles emitted at small angles are also
detected by a scintillator wall, that is installed
three meters from the target and permits
particle identification by means of their time
of flight and their energy loss in the scintillators.
The particle identification in the central region
is accomplished with a plastic scintillator barrel through
the measurement of dE/dx.
\begin{figure}[h]
\centerline{\epsfig{file=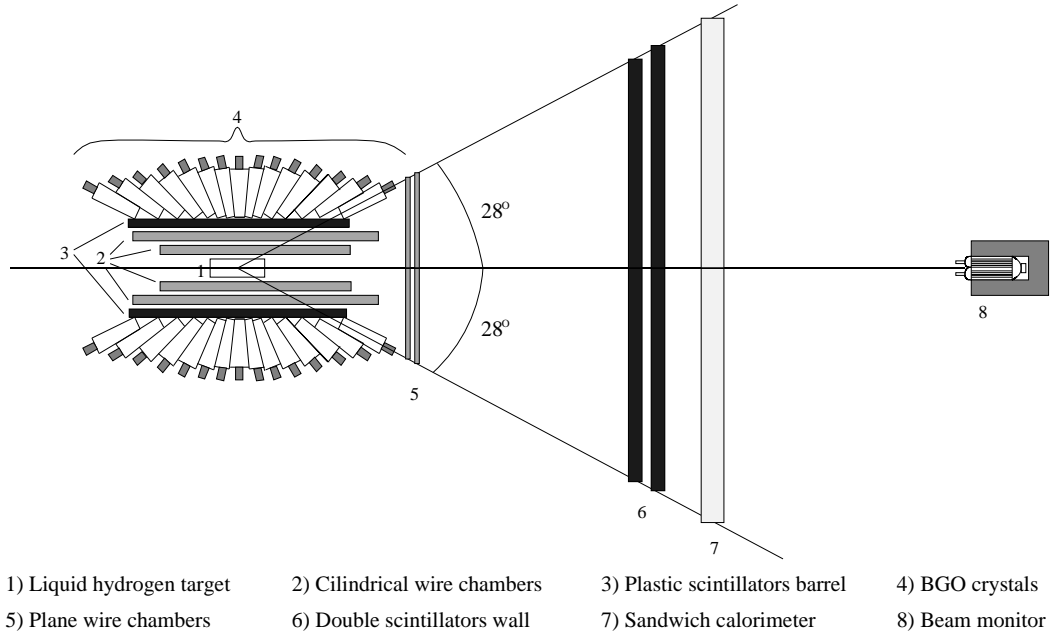,width=5.5truein}}
\caption{Layout of the GRAAL apparatus.}
\label{graal}
\end{figure}
\par In this paper we describe the photon-detection 
system, which has been designed to measure
the energy of $\gamma$ and neutral mesons
decaying in two or more photons
($\pi^{0}$, $\eta$, $\eta'$, K$^{0}$, $\phi$), with a good
angular resolution.
The calorimeter covers 90\% of the entire solid angle,
detecting particles emitted at angles from 25\gradi to 155\gradi.
\par The accuracy and reliability of the energy calibration 
is a basic requirement
for this detector in which a large number of BGO sectors, comprising
about 500 PMs, are involved and high resolution is expected.
The problem is to keep
under control the variations in the gain and temperature
of the different sectors as a function of time,
thus ensuring uniformity of response during
data taking and keeping to a minimum the time
spent calibrating the calorimeter.
We shall, therefore, give particular emphasis
to the description of our LED-based monitoring
system, which plays a key-role in this respect.
\par In sect. 2 we describe briefly the
characteristics of the apparatus.
In sect. 3 the principles
of the electronics and data acquisition.
Sect. 4 is devoted to the calibration 
procedure.
The gain monitoring system is described in sect. 5.
In sect. 6 we report on the linearity
of the calorimeter energy response.
In sect. 7 we report on the performances
of the BGO calorimeter and the monitoring
system, with special emphasis on the energy
resolution and time stability.

\section{Description of the detector}

The BGO detector is shown in detail in fig. 2. The mechanical support
structure consists of 24 baskets of Carbon fiber composite material supported
by an external steel frame. Each basket is divided into 20 cells with very thin
walls, 0.38 mm for the inner and 0.54 mm for the outer walls, to keep the
crystals optically and mechanically separated.
The Carbon fiber has been preferred to other materials like Aluminum for its
higher rigidity and lower gamma ray attenuation due to its low Z number.
The support frame is divided into
two halves which can be taken apart by 1.5 meters to allow access to the target
and central detector region. When closed the structure leaves a 20 cm
diameter hole
along the beam-line for the insertion of the target, the cylindrical wire
chambers and the plastic scintillator barrel.
\vspace{0.45truein}
\begin{figure}[h]
\centerline{\epsfig{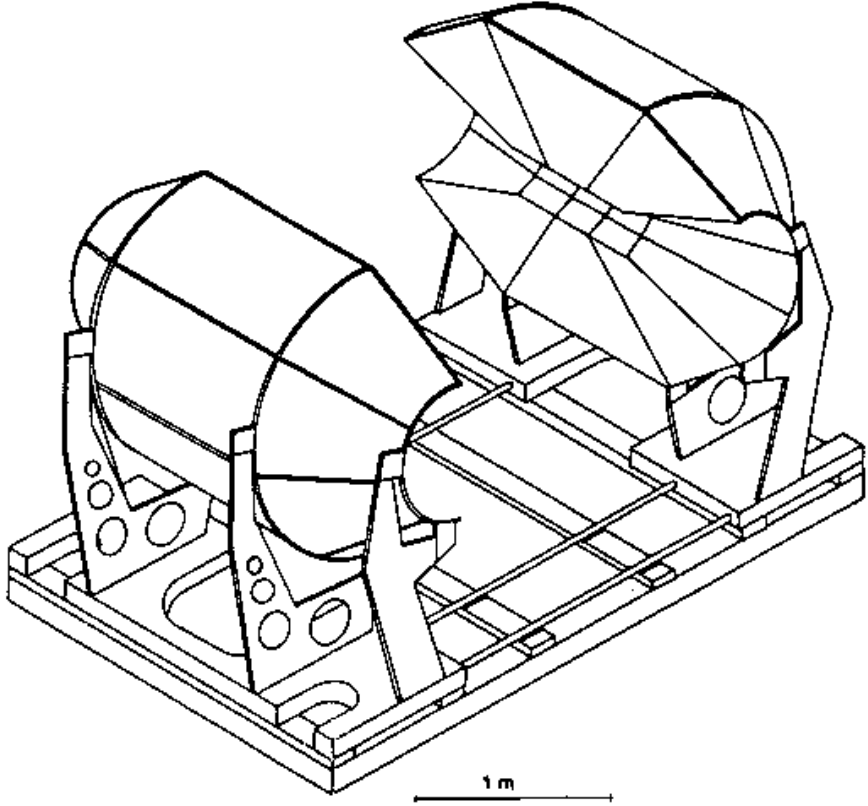}}
\caption{A sketch of the BGO calorimeter showing the carbon fiber baskets 
mounted on the external support frame separable into two halves.}
\label{bgo}
\end{figure}

\par The crystals are of 8 different dimensions and are
shaped like pyramidal sectors with trapezoidal basis \cite{cryst} (see fig.3).
They define 15 angular regions ($\vartheta$) in the plane containing the 
symmetry axis of the calorimeter, coincident with the beam axis, 
and 32 ($\varphi$) in the plane orthogonal to 
the beam axis (see Tab.1 of \cite{bgo} and \cite{zuc92} for details).
The 480 crystals have all the same length of 24 cm ($>$21 radiation lengths),
for a good confinement of photon showers in the GeV region, and are arranged
in such a way that the 
reaction products emitted
in all directions from the target center
encounter a constant thickness of BGO.
Each crystal is wrapped up in a thin (30$\mu$m) aluminized mylar reflector,
and its back side is optically coupled to a photomultiplier (PM) \cite{hama}.
Two holes in the back side of the crystal support are used for 
monitoring the temperature and for the input,  
through optical fiber, of light pulses  which are used for the measurements
of the linearity and gain stability of the photomultipliers.\\

\begin{figure}[h]
\centerline{\epsfig{file=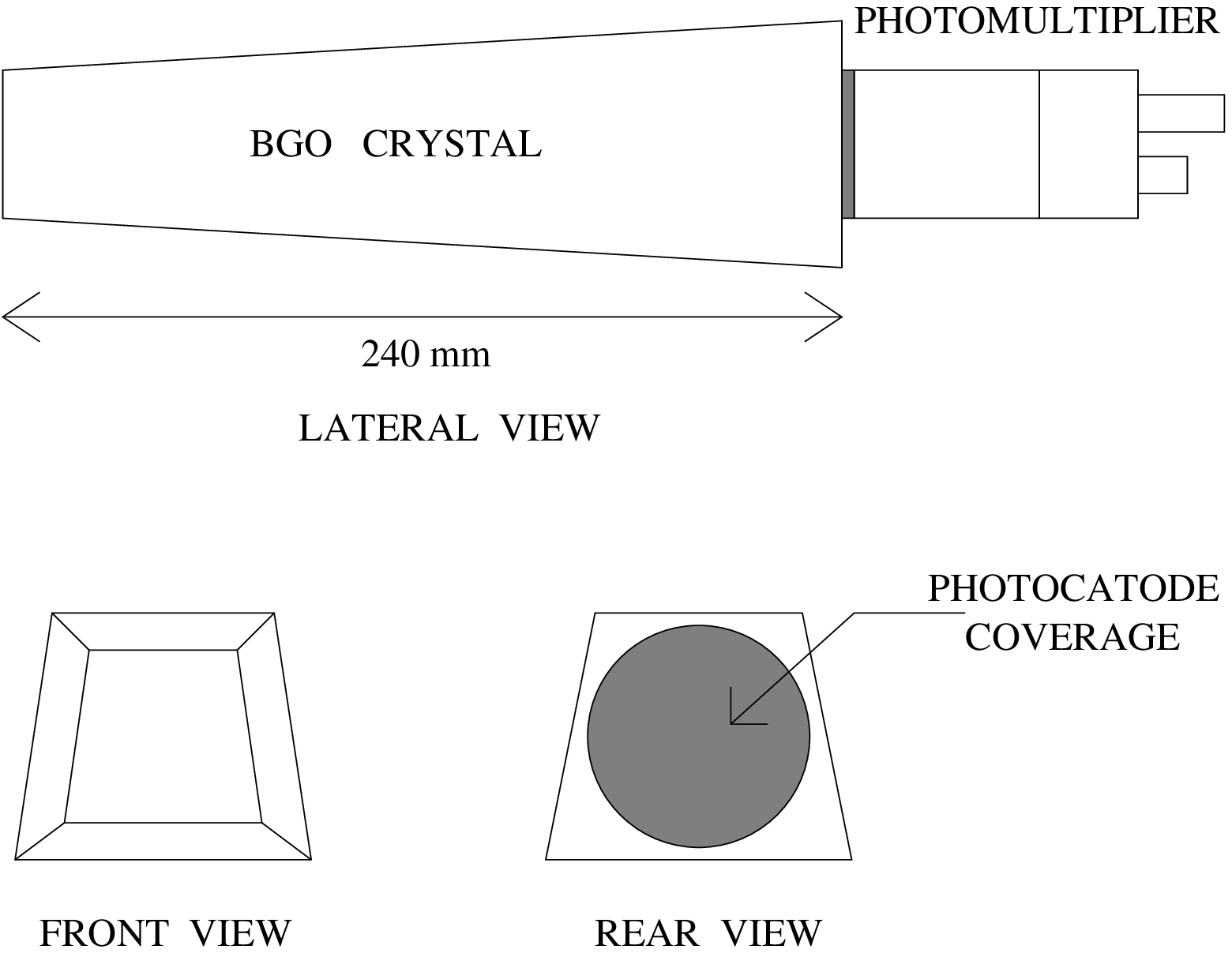,width=4.5truein}}
\caption{Schematic views of a BGO crystal.}
\label{crist}
\end{figure}

During the production phase each one of the crystals has been accurately
tested to check the accomplishment of the requirements imposed for
acceptance: longitudinal uniformity $\geq$ 95\% and resolution at 
the 0.661 MeV Cesium $\gamma$-peak $\leq$ 20\% FWHM. 
The quality tests gave results better
than the design specifications. Two thirds of the crystals have an average
resolution at Cesium better than 18\% FWHM and two thirds have a longitudinal 
uniformity greater than 97\%.

\section{Electronics and data acquisition}

\subsection{The linear chain}
Since the BGO calorimeter is operating in a region 
without magnetic field and
we need to measure with a good resolution electromagnetic
showers with energy less than few
hundreds MeV, we choose for the readout of the signal standard photomultipliers
due to their noise much smaller than that of other devices such as photodiodes.

The anode signals from the PMs enter in 15 adders (MIXER), 
each having 32 input channels with programmable
attenuators  \cite{caen}. 
The outputs from each module consist of :
a linearly summed prompt output, with a fan-out of 6,
used for trigger purposes and to build up the calorimeter total energy
hardware sum.
A 300 ns delayed and, if necessary, 
attenuated output that is sent for digitization to two
FERA modules (Fast Encoding and Readout Adc, charge-sensitive, 11-bit,
16 channels per unit) \cite{Lecro}.
The linearly summed output of each MIXER ($\Sigma_{\vartheta}$) corresponds 
to the sum of the 
signals coming from the 32 BGO crystals having the same $\vartheta$ angle.
The 15 $\Sigma_{\vartheta}$ outputs are sent to another MIXER (MIXER-$\Sigma$).
The individual linear output signals from the MIXER-$\Sigma$ 
are sent to another FERA ADC
for the acquisition of the partial sums. The summed output signal from the
MIXER-$\Sigma$, that corresponds to the total energy of the
BGO calorimeter, is sent to the different trigger circuits 
and to another FERA channel (see fig. 4).
The linear part of the acquisition electronics is thus formed by 16 MIXER
and 31 FERA ADC modules.\\

\begin{figure}[h]
\centerline{\epsfig{file=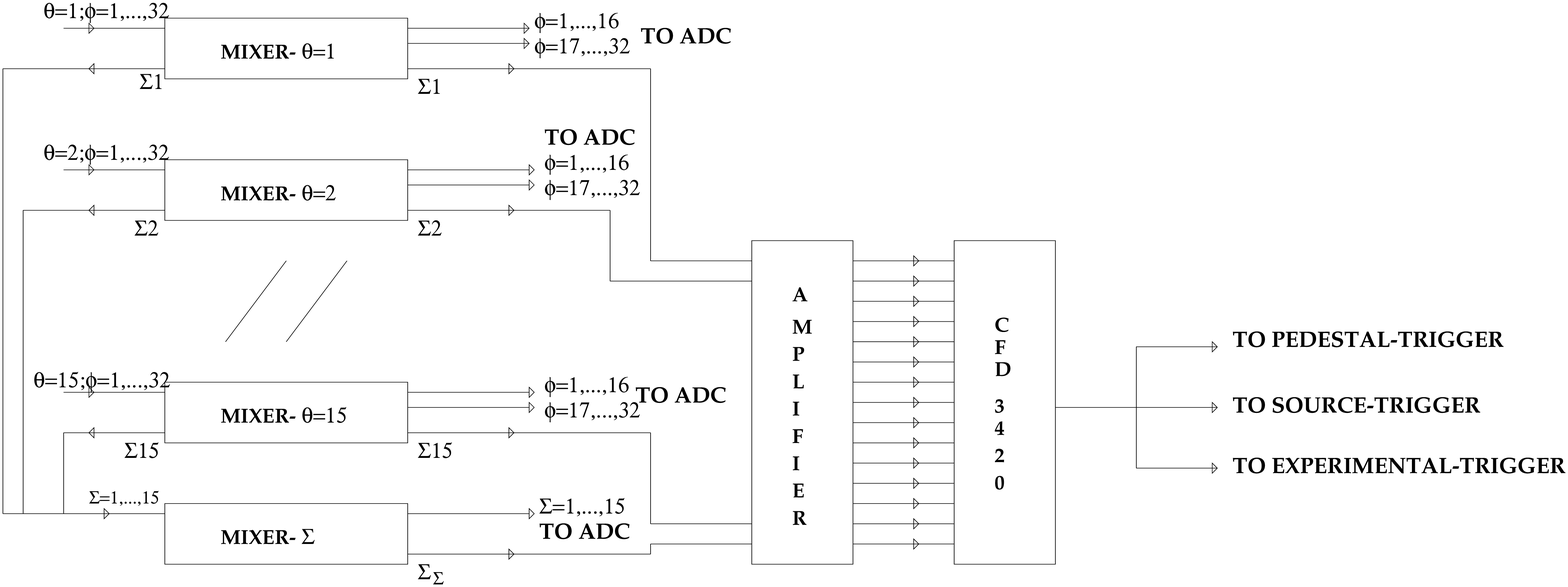,width=5.5truein}}
\caption{Scheme of the linear chain of the electronics.}
\label{mixer}
\end{figure}

\subsection{The readout system}
For the ADC read-out of the BGO calorimeter and other detectors of the
GRAAL apparatus we have found an original solution that allow a fast and
sequential acquisition of all the CAMAC Crates using the FERA-BUS.
In fig. 5 is shown the scheme of the handshake logic for the read-out 
of the ADC.
The FERA ADC of each crate are connected in parallel to the FERA-DRIVER
\cite{Lecro1}
through the COMMAND-BUS and the DATA-BUS. We have specifically designed
a system that needs only one BUS for the data coming from the
DRIVERS of each Crate. For this purpose we have used three BUS-SWITCH modules
\cite{caen1} as interfaces between the FERA-DRIVERS 
and the single DATA-BUS. The
BUS-SWITCH module was originally designed to accept 
as input the signals from two
different DATA-BUS giving at the output one of the two depending on the status 
of the strobe signals of the two DATA-BUS. In our scheme we exploit the
three state port characteristic of the BUS-SWITCH output. 
The open circuit state of
the output allows to disconnect the module from the BUS, in such a way
that the transmission on the same BUS of signals
from other modules is not disturbed.
Using one BUS-SWITCH module for each crate and connecting 
the DRIVER data output
with one of the two BUS-SWITCH data inputs, 
with a suitable signal timing logic,
we achieve the "mixing" of the data on a single BUS.
The FERA-ADC are sequentially read starting from the third Crate and continuing
with the ADC of the second and first 
Crate pertaining to the BGO calorimeter.\\

\begin{figure}[h]
\centerline{\epsfig{file=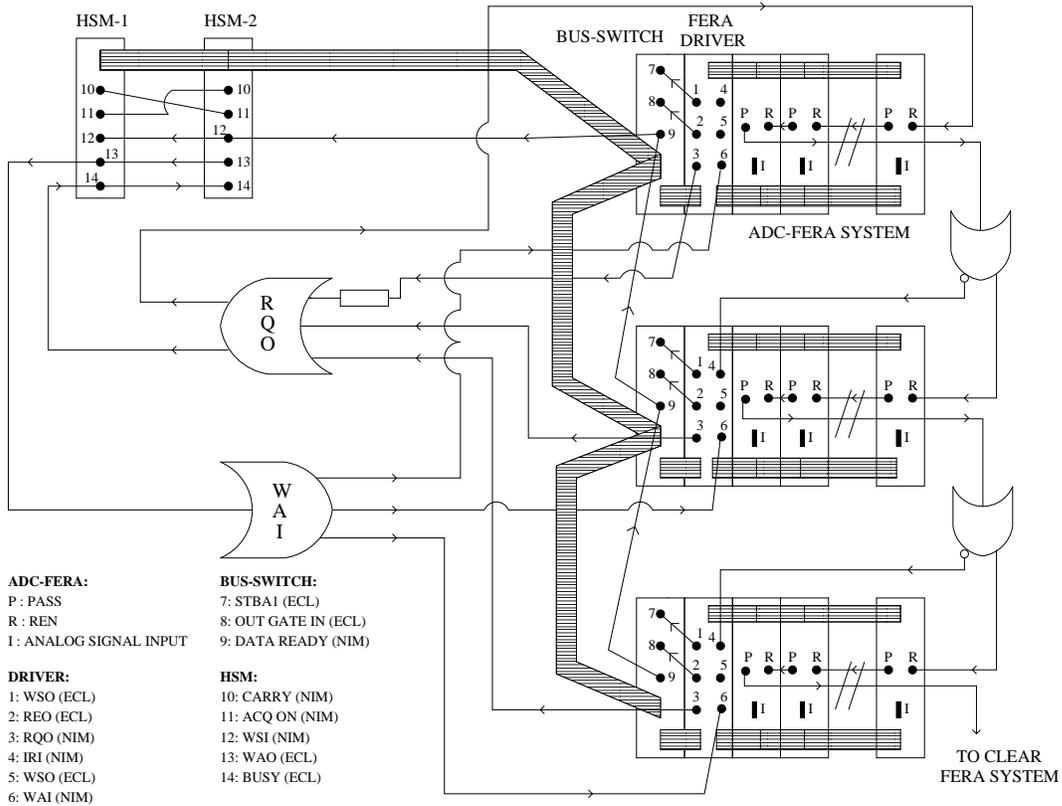,width=5.5truein}}
\caption{Scheme of the read-out system and handshake logic.}
\label{hand}
\end{figure}

During the calibration phase the data are temporarily stored in an auxiliary
memory area formed by two High Speed Memory modules \cite{CES} HSM-1 and HSM-2,
as shown in fig.5. These memories are configured to work in 
a flip-flop to combine
acquisition and data readout without loss of speed. When HSM-2 is filling,
HSM-1 transfers the recorded data to the memory of the calibration
processor \cite{CES1}, according to an Interrupt scheme, and will be ready to
accept new data when HSM-2 is full.

During the experimental data acquisition the gate signal 
for the FERA-ADC is given by
the experimental trigger and the FERA-BUS is automatically connected to the
general data-bus for the global acquisition of all the detectors of the
GRAAL apparatus.

\section{The calibration and equalization method}

The absolute calibration of the crystals is obtained
using the 1.27 MeV photons ($E_{source}$) from a $^{22}$Na source. 
The response of the 480 BGO sectors is equalized 
using an automatic procedure, that sets all the
PM's gains, varying their high voltage. 
If we call $c_{eq}$ the calibration channel
we would like to obtain, we can say that 
the equalization is achieved when all the ADC channels
$c_{i}$ ($i=1,\ldots,480$) are within
the following range:

\[
\frac{| c_{i}-c_{eq}|}{c_{eq}}\le B(\%)
\]

\noindent $B(\%)$ indicates the precision of the procedure.
To maintain the time spent for the equalization at the reasonable
value of about one hour, we fix $B=1.5\%$.
This value affects only the precision on the 
hardware energy sum used for the experimental trigger, 
but does not affect the absolute calibration resolution because
in the data analysis the exact value
of the calibration constant for each crystal is used.  

The choice of the equalization channel is determined
by the maximum energy value that has to be measured
and by the range of the detector linearity response.
The maximum $\gamma$ energy of the GRAAL beam is 1.5 GeV,
but the Monte Carlo simulations and the experimental data
\cite{Bonn}
indicate that no more than 80\% of the incident
$\gamma$ energy can be contained in a single crystal 
($E_{max}=1.2$ GeV),
due to the transverse spread of the electromagnetic shower.
The saturation effects of the crystals PMs become evident
for peak pulse amplitudes of the order of 5 V, so that 
the signals corresponding to the maximum energy released
should not exceed this value. For the signals coming from
BGO scintillation events, we have measured the following relation
between the peak amplitude of the signal $V_{peak}$(mV) and
the ADC channels:
\[
V_{peak}=10 \mbox{ mV } \Rightarrow~~100 \mbox{ channels}
\] 

The end-scale value of our FERA-ADC is 1920 channels, corresponding
to a signal equivalent charge of 480 pC. To this value
we must subtract the channels corresponding to the pedestal
signal, that have been adjusted in the range 100-120;
for this reason we can assume that the maximum number of
ADC channels available is: 
$c_{e-s}= 1800$. 
In the hypothesis of a linear response, the following relation
holds:
\begin{equation}
E_{max} = \frac{c_{e-s}\cdot E_{source}}{c_{eq}}\cdot \frac{1}{f}
\label{emax}
\end{equation}

\noindent where $f$ is the
attenuation factor on the analog signal introduced
by the MIXER. According to (\ref{emax}), we obtain in our case:
\begin{equation}
c_{eq}\cdot f=1.91
\label{ceq}
\end{equation}

\noindent The best value for $f$ is determined
knowing that the $V_{peak}$ values of the signal corresponding
to $c_{e-s}$ depend on the attenuation factor, according to
the following relation:
\[
V_{peak}(\mbox{mV}) = \frac{c_{e-s}}{f}
\cdot 0.1 \mbox{ mV}/\mbox{channel} = \frac{180 \mbox{ mV}}{f}
\] 

\noindent from which we decided to use $f=1/32$, that
represents the best approximation
between the available attenuation factors.
From eq.(\ref{ceq}) it follows that $c_{eq}$=64
is a reasonable compromise between calibration
precision and saturation effects.
This way, without attenuation, we obtain 
a calibration constant $m_{eq}\simeq$0.02 MeV/channel.
The maximum energy that can be measured with
attenuation is thus
equal to $\simeq$1.15 GeV compatible with the expected $E_{max}$ value.\\

\section{The calibration monitoring system}

\subsection{Introduction}

The calibration constant of each crystal
measured at the equalization time $t_{0}$ 
may change as a function of time, due to
two principal reasons: variation in the 
crystal light output due to temperature effects and
variation in the PM gain.
\par To control the temperature effects we have set up 
a monitoring system, that is
discussed in details in \cite{Rachele}
and will not be described further here.

Gain variations of the photomultipliers may occur
as a consequence of the following effects:

\begin{itemize}
\item photocathode temperature variation: 
at the BGO peak emission wavelength (480 nm) the efficiency variation 
of a bialkali photocathode is of the order of -0.4\%/$^{\circ}$C;
\item instability in the high voltage power supply \cite{caen2}:
these are of the order of $\pm$1 V and assuming a HV typical value
V=1500 V, the induced gain variations are 0.6\% for 
the R580 PM model  
and 0.8\% for the R329-02 model; during the acquisition of the
calibration peaks, this effect is however strongly reduced
as it is averaged over a long time interval;
\item PM drifts and shifts; for both source and experimental
events, we have an anodic mean current of the order of $10^{-2}$ $\mu$A,
much lower than the maximum value of 1 $\mu$A recommended
in order to avoid drift effects; also shift effects 
can be neglected as previous tests made on a
basket of 20 BGO crystals \cite{Bonn} have shown that
for rates lower than 1 KHz, as in our case, we have
a gain variation due to shift effects of the order of 0.01\%;  
\item aging of cathode and dynodes materials: 
a typical time scale for these effects is of the order of
few years and can be kept under control with
a regular monitoring of the PMs gains;
\item voltage dividers instability: this effect does not cause a time
variation of the PM gain but a loss of linearity in the PM response;
the linearity curves have been measured
as reported in section 6.
\end{itemize}
\par In order to keep under control all time variations
of the calibration constants previously discussed
we have realized a photomultiplier gain monitoring system
as described below.\\

\begin{figure}[h]
\centerline{\epsfig{file=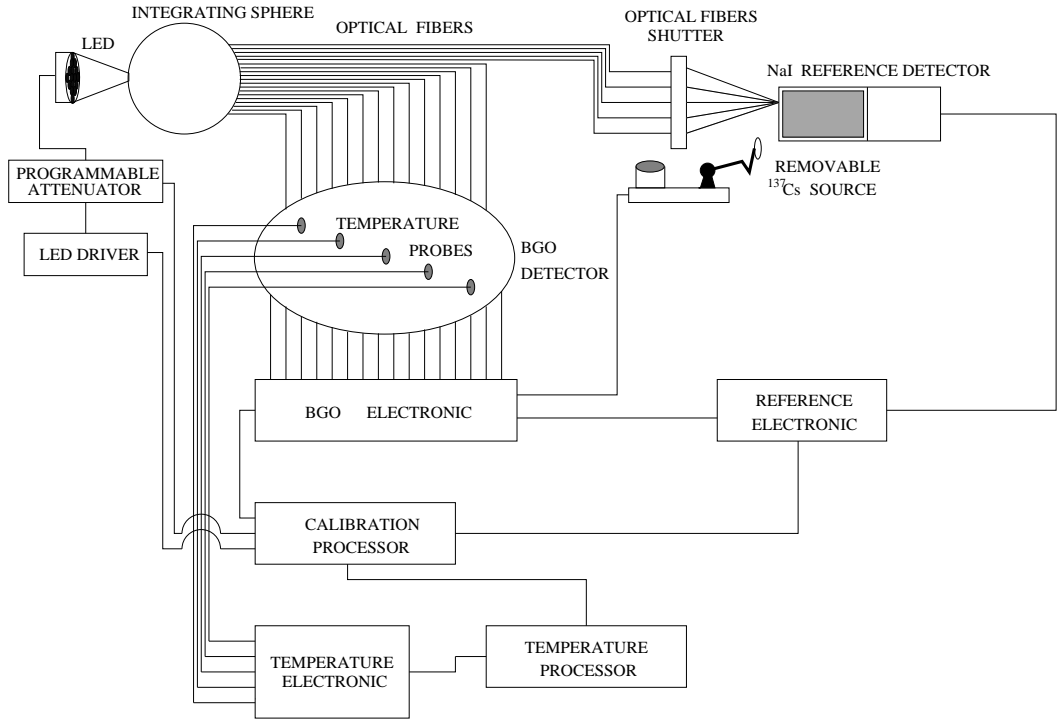,width=5.5truein}}
\caption{Schematic view of the calorimeter gain and temperature monitoring
	 system.}
\label{strum}
\end{figure}

\subsection{The photomultiplier gain monitoring system}
 
The monitoring system, as shown in fig. 6, is made of:
\begin{itemize}
\item high luminosity, blue emitting, pulsed LEDs as light source; 
\item a light distribution system consisting of an integrating 
sphere and optical fibers;
\item a reference NaI scintillation detector.
\end{itemize}
 
 A matrix of 7 LEDs \cite{led}, is directly coupled to the input port of the
 integrating sphere. 
 The following aspects were taken into account with
particular care in the final choice of the light source:
\begin{itemize}
\item emission spectrum;
\item time stability;
\item time shape of the light pulse;
\item intensity.
\end{itemize} 

The emission spectrum of the chosen LEDs has a maximum 
at 450 nm with a FWHM of 140 nm, that is quite similar
to the BGO one.

The LED pulse-to-pulse stability is of the order 
of 1\% , depending on the light intensity.
In fig. 7 is shown the LED pulse distribution   
corresponding to the maximum light intensity, in a generic BGO crystal. 
Measurements done in 
a time interval of several days
have shown that the LED time stability is of the order of
a few percent.

\begin{figure}[h]
\centerline{\epsfig{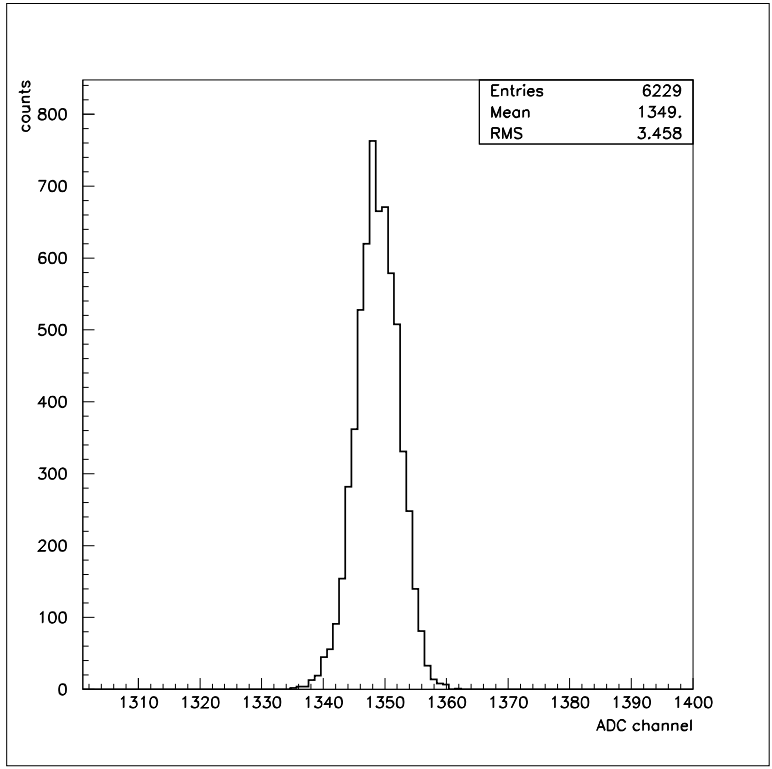}}
\caption{LEDs pulse to pulse stability distribution.}
\label{ledstab}
\end{figure}

We decided not to use the pulsed laser source \cite{laser},
in spite of its high intensity, because it had 
a pulse to pulse instability greater than 20\% (FWHM)
and showed a time drift in the emission intensity 
as high as 50\%.
\par The time shape of the light pulse is important for two fundamental
reasons. The first is that the PM anode pulse from the light source  
should reach a charge equivalent to $\sim$1 GeV energy without having a
current peak that gives rise to saturation in the PM.
The second reason is that the PM response may depend on the time shape of the
input pulse also without saturation effects.
Especially for these reasons we realized a driver circuit for the LEDs that
produces an anode pulse, see fig. 8, very similar to that produced by 
scintillation in BGO.

The intensity of the light source should be high enough to simulate the 
maximum
energy deposited in each detector. The actual configuration of the LEDs
produce an equivalent energy of almost 1 GeV for each of the 480 BGO
crystals.
\vspace{0.2truein}
\begin{figure}[h]
\centerline{\epsfig{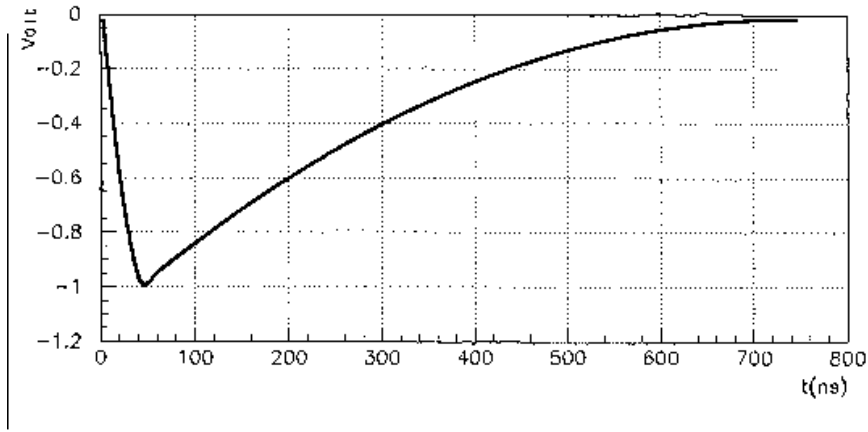}}
\caption{Anodic pulse shape generated by the LEDs driver circuit.}
\label{led}
\end{figure}

We have chosen an integrating sphere of 6" internal diameter \cite{sphere} 
to match the LEDs light to the 600 optical
fibers, 6 meters long, carrying the light to each detector. 
This solves in a simple way the
challenging problem of having a uniform light distribution over the fibers
that are connected to the exit port of the sphere. Thus the ratios of the 
incident light intensities over any couple of fibers are constants independently
of the variations of the light source.
The inner surface of the sphere consists of Spectralon reflectance material
that gives the highest diffuse reflectance of any known material over the
UV-VIS-NIR region of the spectrum. The reflectance is generally $>$99\% over the
range 400-1500 nm.
Radiation introduced into the sphere undergoes multiple scattering, which
results in a uniform, spatially integrated radiation distribution
within the sphere. The exit port serves as a source of radiation characterized
by a radiance which is constant over the plane of the exit port and independent
of viewing angle.
The light carried by each fiber is then optically coupled to each crystal.
\par The reference detector is a crucial part of the linearity and gain 
monitoring
system because it measures the intensity of the light pulses from the LEDs and
the outputs of
all the other detectors are compared to this value. Thus its linearity and gain
variation should always be under control.
This detector is a 2"x2" NAI(Tl) scintillator \cite{nai} coupled to the same 
type of PM used for the bigger BGO crystals. Five optical fibers carrying the
LEDs light are coupled to the detector through a quartz window.
A $^{137}$Cs radioactive source was used to monitor the gain drift of its
response. We have realized a system to remotely remove the source
when the LEDs light is flashing to avoid that random coincidences between the
LEDs and the source signal alter the measure of the LEDs
intensity.

The electronic chain and trigger circuit for the acquisition of the reference 
detector are different from those of the calorimeter. 
The anode signal is formed
by a preamplifier and a spectroscopy amplifier and is then digitized by a High
Performance Buffered Spectroscopy ADC \cite{adc} with 8000 channels.
The data from the ADC are sent via an external bus to an Histogramming Memory
\cite{adc} and at the end of the acquisition are transfered to the  
calibration processor memory for analysis.

\section{Correction of non-linearities}
 
 Our method of calibration of the BGO calorimeter, as described in section 
 4, is simple, fast and very effective, allowing to determine the absolute
 calibration constants of all the cristals in less than ten minutes during a
 pause of the experimental data taking, tipically a new injection of the
 electron beam in the ESRF. Nevertheless the extrapolation of the calibration
 constant from the 1 MeV region of the gamma source to the 1 GeV region of the
 maximum energy deposited in a single cristal requires careful 
 account of the possible non-linear behaviour of the electronics chain (PMs,
 MIXER and FERA ADC) in this large energy range.
 We have thus measured the linearity curve of each crystal of the calorimeter
 and applied these correction factors to the measured energy.
 
 The variations of the light intensity of the LEDs over almost 3 orders of
 magnitude was obtained attenuating the driving LEDs input pulse using a CAMAC
 programmable attenuator \cite{caen3} with steps of 0.25 dB.
 The relative intensities of the light pulses for each crystal were normalized 
 to the response of the reference detector, the excellent linearity of which
 has been repeatedly verified.
\begin{figure}[h]
\centerline{\epsfig{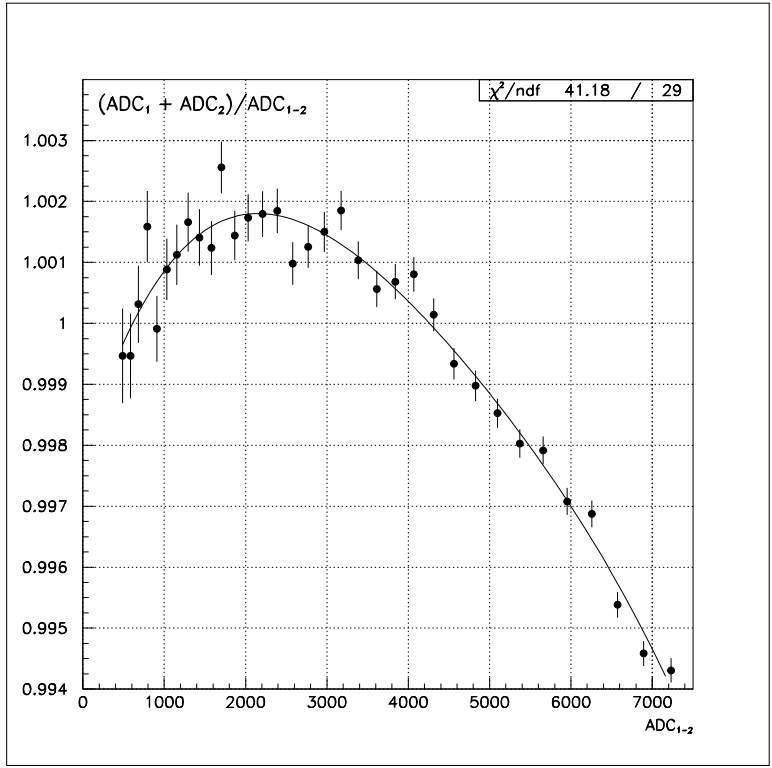}}
\caption{Ratio between the sum of the ADC channels corresponding
 	to the two single fibers (ADC$_1$+ADC$_2$)
 	and the channel obtained when the shutters of the two fibers were both
 	opened (ADC$_{1-2}$).}
\label{lin}
\end{figure}

 \subsection{The measurement of the reference detector linearity}
 
 The measure of the NaI reference detector linearity is crucial and 
 preliminary to that of the BGO detectors.
 We set up a system that allows to send to the reference the light of any 
 combination of five fibers. 
 The five fibers coming from the integrating sphere are in fact coupled to
 other five fibers carrying the light to the reference in such a way that it is
 possible to adjust the distance between each pair of coupled fibers and thus 
 adjust, for each fiber, the light reaching the detector. Moreover a remotely
 controlled shutter placed between each couple of fibers 
 enables or disables the optical contact.
\begin{figure}[h]
\centerline{\epsfig{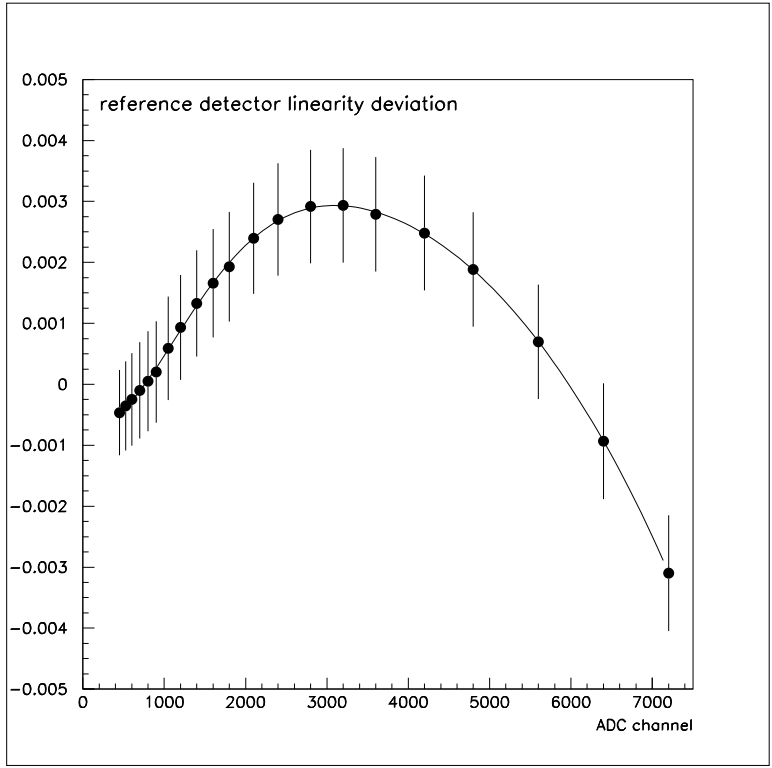}}
\caption{The integral non-linearity distribution of the reference detector
	respect to the channel 200.}
\label{fit}
\end{figure}

 \par Two out of five couples of fibers are adjusted 
 in such a way that they send the same amount of light to the
 reference.
 The output of the reference detector is measured three times : when one shutter
 is open, when the other one is open and when both of them are open. 
 If the system is linear the sum of the individual measurements should equal, 
 for any light intensity, the response when
 both shutters are open. 
 The variations of the LED light intensity during the three measurements, 
 of the order of few tenths of
 percent, are monitored and corrected using the light 
 measured by one BGO detector.
 In fig. 9 it is shown the ratio between the sum of the ADC channels 
 corresponding to the two single fibers
 and the channel obtained when the shutters of the two fibers were both
 opened. Each point represents the linearity of that ADC channel respect to 
 its half. 
 These points are then fitted with a polynomial as shown in the same figure.
 From this curve, moving in steps of a factor of two, we can deduce the integral
 non linearity points of fig. 10.
 This figure indicates that the non linearity does not exceed $\pm$0.3\% in a
 range larger than a factor of 30.
 These data are fitted again with a polynomial function to obtain the curve
 shown in fig. 10. 
 Applying this correction to the original points of fig. 9 we obtain
 the distribution of fig. 11 which indicates that, after correction, the
 distribution of the individual measurements is better than $\pm$0.1\%.
\begin{figure}[h]
\centerline{\epsfig{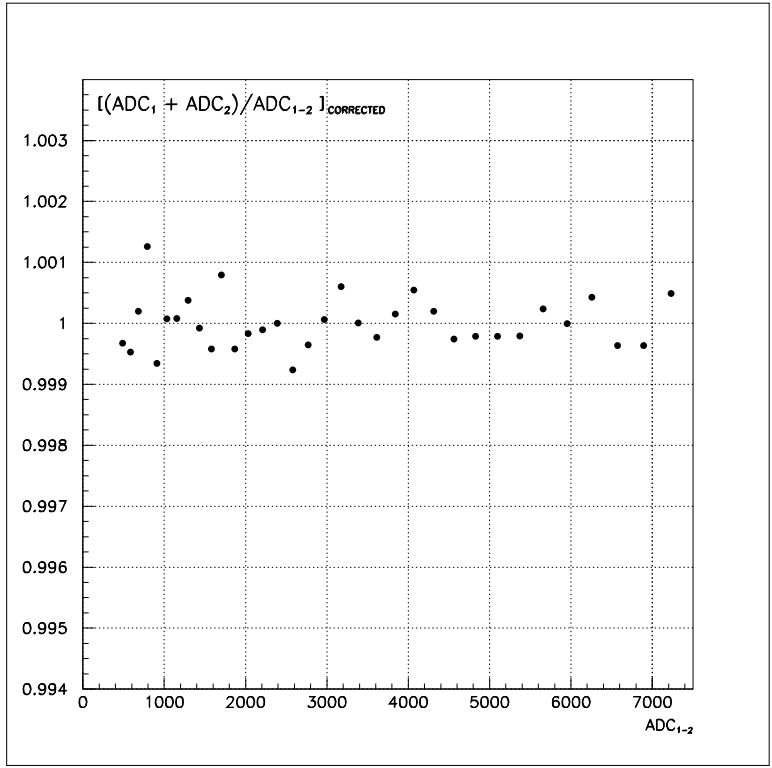}}
\caption{Same distribution of fig. 9 after the correction of the points with the
	non-linearity curve of fig. 10.}
\label{cor}
\end{figure}

 With this procedure we can measure the linearity of the BGO over a factor of 30
 extending approximately from 1 to 30 MeV.
 To extend our measurements of a further factor of 30 to arrive, 
 in the BGO, at a
 light pulse equivalent to around 1 GeV, we shut off four of the five optical
 fibers delivering the light pulse to the reference detector. The only fiber
 left open had been previously adjusted to carry about 1/30 of the total
 light. This way the pulse in the reference detector returns to around
 channel 200. Increasing further the light of another factor of 30, up to its
 maximum value, we can extend our measurements over three orders of magnitude. 
 This way the absolute calibration done with the $^{22}$Na source at 1.27 MeV
 is extended up to 1 GeV.
 The result is a calibrated light source which covers a range of 1000:1.
 
 \subsection{The measurement of the BGO detectors linearity}
 
 In order to correct the non linear response of the BGO detectors, we developed
 an automatic procedure to build the linearity curves for each crystal.
 The LEDs light intensity has been initially adjusted in such a way 
 to correspond to an
 energy of $\sim$1.27 MeV around the ADC calibration channel 64, with no
 attenuation. Successively a 30 dB attenuation
 on the analog signal is inserted and the LEDs light is progressively
 increased until the maximum intensity corresponds to an equivalent energy of
 almost 1 GeV. 
\begin{figure}[h]
\centerline{\epsfig{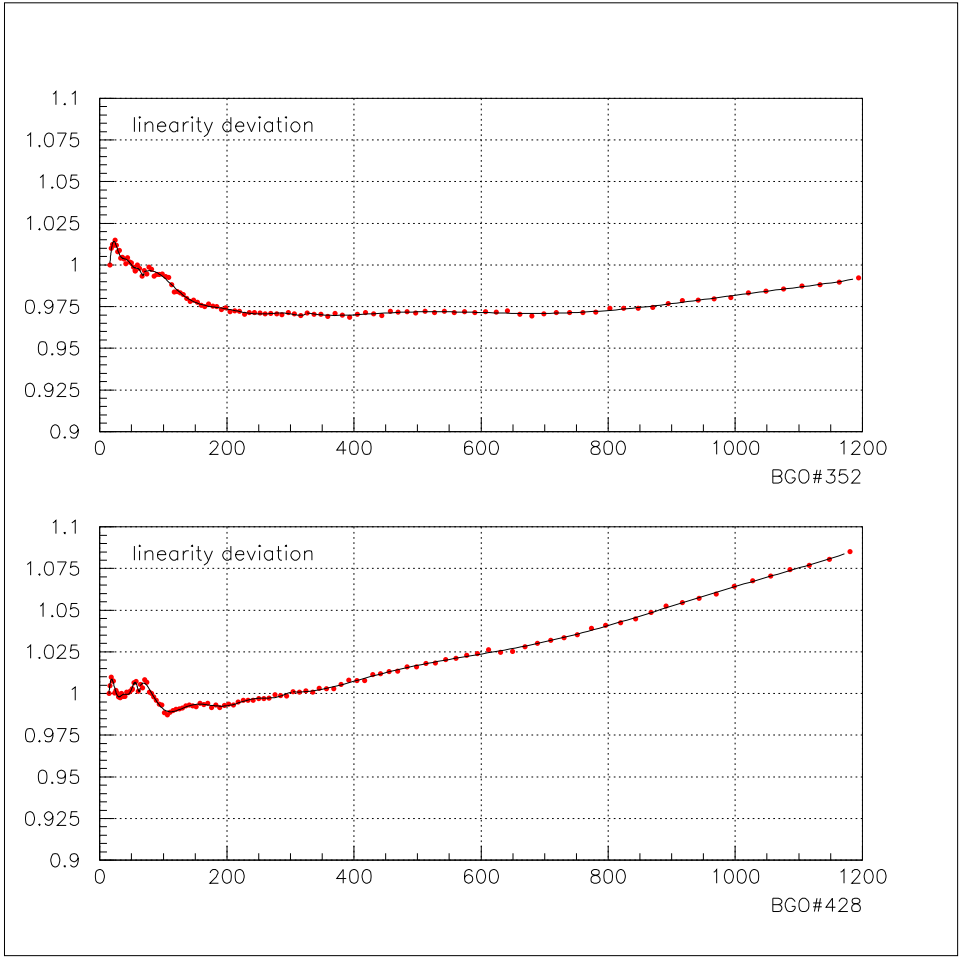}}
\caption{Two typical linearity deviation distribution corresponding 
to the two types of photomultipliers used for the BGO calorimeter.}
\label{bgolin}
\end{figure}

 This way we obtain the response of each BGO detector to light pulses of
 known amplitude corresponding to energies up to 1 GeV.
 Two typical response curves from two photomultipliers of different type are
 indicated in fig. 12. They have been taken with an attenuation of 30 dB.
 Each distribution was separated in three regions and each region 
 fitted with a polynomial function up to the eighth degree. 
 The energy correction factor for each ADC channel of each
 crystal has been extracted from these functions.
 
 The absolute energy calibration of the ADC scale for each BGO detector
 with the 30 dB attenuation
 is obtained calculating the energy corresponding to the first point of the
 linearity distribution at 30 dB. This is done, for each BGO crystal, 
 multiplying the energy
 calibration by the ratio between the light measured in the reference, in
 correspondence to this point at 30 dB, and that corresponding to the
 calibration channel at 0 dB.
 
  \section{The BGO calorimeter performances: energy resolution and time 
  stability}
  
  The validity of the calibration method and the linearity correction
  procedure for the BGO calorimeter have been recently tested using  the 
  first experimental data obtained with the polarized and tagged 
  GRAAL $\gamma$ ray beam
  incident on a liquid hydrogen target placed at the center of the calorimeter.
\begin{figure}[h]
\centerline{\epsfig{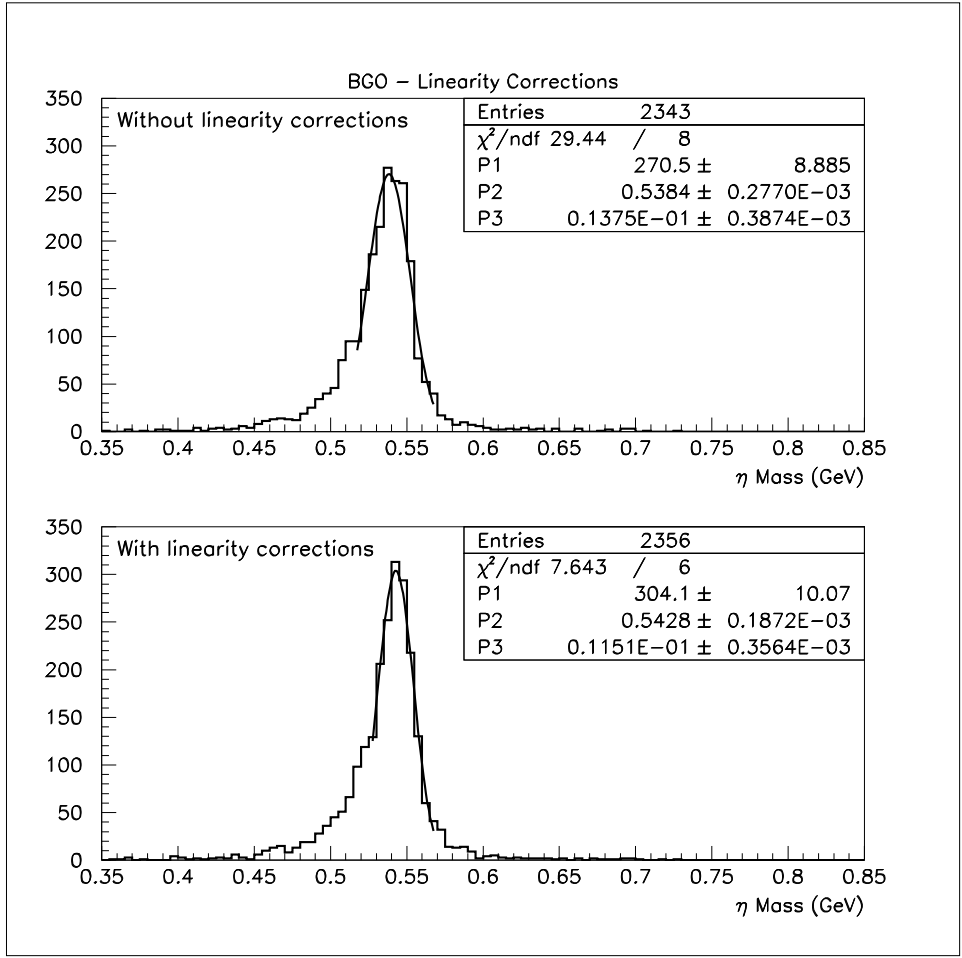}}
\caption{Reconstructed $\eta$ mass distribution with and without the BGO
	linearity corrections. P1, P2, P3 are the parameters of the gaussian
	fit.}
\label{eta}
\end{figure}

  In particular we looked at the reaction $\gamma + p \rightarrow \eta + p.$
  In order to identify the $\eta$ mesons we selected the events with two neutral
  clusters in the calorimeter, corresponding to the decay $\eta \rightarrow
  2\gamma$. We required also no charged cluster in the calorimeter and one
  charged particle in the forward scintillator wall, corresponding to the
  recoil proton. In fig. 13a and 13b are shown the reconstructed $\eta$ mass
  distribution with and without the BGO linearity corrections. The value of the
  $\eta$ mass has been deduced using the energy of the two neutral clusters, the
  incident $\gamma$ energy from the tagging system and the angle of the proton
  in the forward scintillator wall.

  The two distributions in fig. 13 have been fitted with two gaussian 
  functions and their
  parameters are reported in the figures, where P2 corresponds to the mean value
  and P3 to the $\sigma$ of each gaussian.
  
  From these figures we can see that the linearity corrections increase the
  $\eta$ mass by $\sim$ 1\% and improve the resolution
  of the mass distribution by almost 20\%.
  
  This way the $\eta$ mass distribution has a mean 
  value of 542.8 MeV
  that differs by less than 1\% from the true value (547.45 MeV),
  due to uncertainties
  on the incident $\gamma$ ray energy and on the
  corrections introduced by the Monte Carlo simulation.
  The value of the $\eta$ mass and its resolution, comparable to that
  measured for the BGO at these energies \cite{Bonn},   
  proves the validity of this method of calibration.
\begin{figure}[h]
\centerline{\epsfig{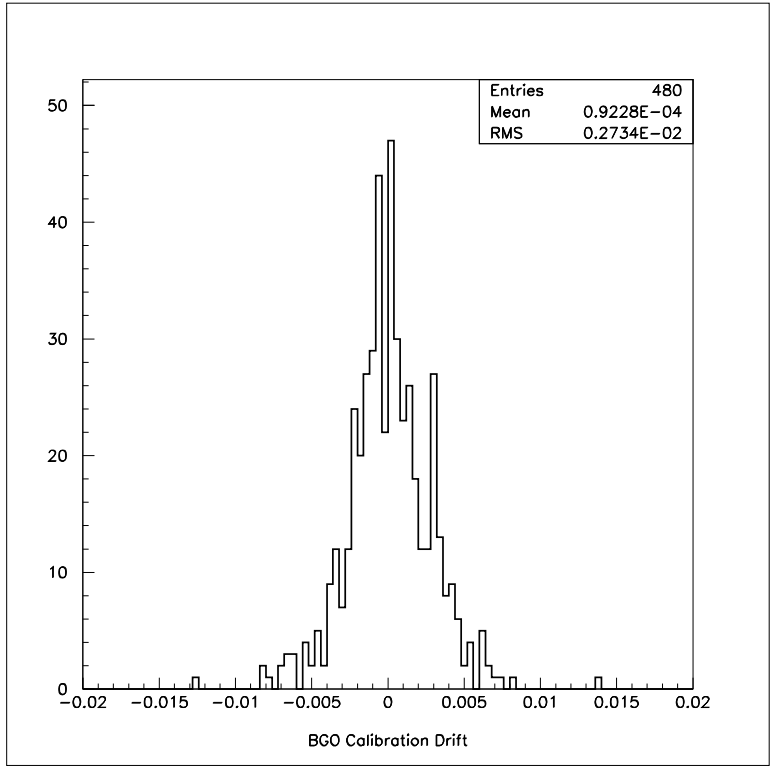}}
\caption{Drift distribution of the calibration peaks for all the BGO detectors in
	a time interval of 12 hours.}
\label{drift}
\end{figure}

  A typical drift distribution of the calibration peaks for all the 
  BGO detectors, corresponding to two consecutive calibration runs,
  is showed in fig.14, where we can see
  that in the time interval of 12 hours the peak positions are
  stable at a few tenths of a percent level.
  \section{Conclusions}
  
  \par We have designed and installed a calibration and monitoring system for
  the BGO electromagnetic calorimeter in operation at the GRAAL beam
  of the ESRF facility. 
  
  The calibration of the entire calorimeter is done
  in less than ten minutes during a pause of the experimental data taking,
  tipically twice a day during machine injection. 
  
  We have developed a method for the correction of the
  non-linearity that allows the measure of the energy in a range of three
  orders
  of magnitude with an accuracy that is a small fraction of the intrinsic
  detector resolution. 
  
  The calibration monitoring and the linearity corrections allow to control
  the energy response of the calorimeter for all the data taking period at
  better than one percent.

  \ack{We wish to thank the following technical staff:  
  F. Basti, A. Orlandi, W. Pesci, G. Serafini and 
  A. Viticchi\'e of INFN Frascati 
  National Laboratory; E. Tusi, L. Distante, E. Reali and M. Travaglini
  of University of Roma 'Tor Vergata' for their
  contribution to the realization of the BGO calorimeter, and the Graal
  collaboration for providing the data on the $\eta$ photoproduction.
  
  We are also indebted with M. Albicocco and R. Cardarelli for their precious
  help in designing and constructing prototypes of electronic modules.}


\begin{thebibliography}{99}
\bibitem{hama} 
Photomultiplier tubes R580 (38 mm diameter) for the 192 
smaller crystals and R329-02 (52 mm diameter) for the 288 bigger ones,
both manufactured by Hamamatsu Photonics. These type of PMs have been chosen
after a detailed comparative test with other models; 
see \cite{bgo} for
details.
\bibitem{caen3} CAEN Mod. N147, especially modified for our purposes
\end{thebibliography}
\end{document}